# Towards printed magnetic sensors based on organic diodes

*Sayani Majumdar* [*,1,2], *Himadri S. Majumdar*[1], *Daniel Tobjörk*[1], *and Ronald Österbacka*[1]

[1]Center for Functional Materials and Department of Physics, Åbo Akademi University, Porthansgatan 3, 20500 Turku, Finland

[2]Wihuri Physical Laboratory, Department of Physics and Astronomy, University of Turku, 20014 Turku, Finland

[*] Corresponding author: e-mail - sayani.majumdar@utu.fi, Phone: +358 2 333 5917, Fax: +358 2 2154676

We report the study of magnetotransport properties of regio-regular poly (3-hexyl thiophene) based organic diodes. The devices were fabricated using two different techniques of spin coating and inkjet printing. Positive magnetoresistance (MR) effect was observed at room temperature in all the devices. The highest MR magnitude reached up to 16% for some spin-coated devices and up to 10% in inkjet printed devices. The MR magnitude and line shapes were found to depend strongly on the measuring current. We observed deviation from the theoretically predicted Lorentzian or non-Lorentzian line shape of the MR traces, which is discussed in detail in the article. Although, the printed devices exhibit MR response as high as for the spin coated ones, they still need to be optimized in terms of performance and yield for large scale applications as magnetic sensors.

**Introduction**

Conjugated or conductive polymers are subject of great research interest due to their rich chemical and physical properties and also due to potential application in devices made using simple fabrication techniques suitable for large scale applications [1,2]. One such application is the capability of sensing magnetic field. Reports from late nineties [3,4] show that organic semiconductors and polymers can have sizable magnetic field effect due to change in spin state of the carriers. Very little attention was paid in exploring the application possibilities of such effect until in 2004, when Francis et al. [5] demonstrated that organic diodes show 10-15% change in electrical resistance under an applied magnetic field of 10-15 mT at room temperature. The physical understanding of the effect of magnetic field on the charge (spin) dynamics of the devices is still strongly debated. Different models have been proposed based on spin dynamics induced by hyperfine interaction, namely, electron-hole pair mechanism [6], triplet-exciton polaron quenching model [7] and bipolaron model [8]. The two former models explain the organic magnetoresistance (OMAR) as the effect of magnetic field on the singlet - triplet mixing [6] of excitons and scattering of charge carriers by triplet excitons together with quenching of triplets by paramagnetic centers [7]. However, some experimental results shows that OMAR is seen even in systems having almost single type of charge carriers and very limited possibility of exciton formation.

Earlier, we have investigated the magneto-transport response of regio-regular poly 3(hexyl-thiophene) (RRP3HT) diodes and bulk heterojunction solar cells made from RRP3HT:1-(3-methoxycarbonyl)propyl-1-phenyl-[6,6]-methanofullerene (PCBM) blends and clarified that a reduced probability of pair formation of electrons and holes lead to disappearance of MR [10]. Though in this article we will mainly report the observation of OMAR in inkjet-printed diodes, we would also indicate some deviations of the experimental results of OMAR line shapes from the proposed theoretical models.

Ink-jet printing is very attractive for printing organic electronic components due to advantages like low cost, low material consumption, non-contact patterning and roll-to-roll compatibility. Different organic electronics components like organic light-emitting diodes [11], organic photovoltaic cells [12] and organic thin film transistors [13] have already been fabricated and reported using the inkjet technique. Inspite of the inherent advantages there are some common experimental obstacles with this method namely, clogging of the nozzles and formation of uneven films on the substrate. These problems can, however, be prevented by proper formulations of the inks. Clogging of the nozzles can be avoided by using suitable printing parameters and an ink with a suitable surface tension (around 28 mN/m2), a low or moderate viscosity (8-15 mPas) and a high boiling point (>100 °C) solvent.

**Experimental**

The spin coated diode devices used in the experiment have the structure indium tin oxide (ITO)/poly(3,4-ethylenedioxythiophene)-poly (styrenesulphonate) (PEDOT:PSS) / RRP3HT / lithium fluoride (LiF)/ aluminium (Al). The ITO coated glass electrodes were coated with a very thin layer of PEDOT:PSS and annealed at 120 °C for 15 minutes. The π-conjugated polymer RRP3HT (from Plextronics) was spin coated from an ortho-dichlorobenzene (oDCB) solution and annealed at 120 °C for 15 minutes. Finally the LiF layer and the aluminium electrode were vacuum evaporated to complete the device structure.

All inkjet printing was performed on a Dimatix Materials Printer (DMP-2831) in ambient room atmosphere. The structure of the inkjet printed diodes were Al /RRP3HT/ Silver (Ag), where all layers except for the Al were inkjetted. The formulated inks were filtered through 0.45 and 0.2 μm filters before filling the 1.5 ml cartridges (DMC-11610). The printed diodes were manufactured on flexible plastic substrates (PET-505) with pre-patterned (1.5 mm wide) Al contacts. RRP3HT 0.67 wt.% (from Plextronics) was dissolved in toluene and oDCB (2:1) and was used as the organic semiconductor (OS). After printing three layers with a drop spacing of 20 μm, the substrates were heated on a hot plate at 100 °C for 30 min to assure a complete evaporation of the solvents. The silver top electrodes were then ink-jet printed from SunTronic U5603 ink which has a silver content of 20 wt.% and suitable rheological properties for inkjet printing. To achieve a continuously printed (~1 mm wide) area on the low surface energy RRP3HT surface, while heating the substrate to 60 °C, ten layers were printed using a drop spacing of 20 μm. The sample was finally annealed on a hot plate at 120 °C for 20 min to evaporate solvents and surfactants and to slightly sinter the silver nanoparticles, resulting in a silver top electrode with a surface resistivity of less than 1 Ω/cm.

After fabrication, the devices were transferred into a cryostat and placed in between the pole pieces of the electromagnet capable of producing up to 300 mT magnetic field. The resistance of the device was then measured by sending a constant current through the device and measuring the voltage drop across it in a varying magnetic field.

**Results and Discussions**

We observed as high as 16% positive MR response at room temperature in the spin coated diodes, where MR is defined as:

$$MR = \frac{R_B - R_0}{R_0} = \frac{\Delta R_B}{R_0},$$

where $R_B$ is the device resistance under external magnetic field B and $R_0$ is the zero field resistance. Fig. 1 shows the MR response of a typical diode with the structure ITO/PEDOT:PSS/RRP3HT/LiF/Al measured at room temperature and with a current of 100 μA. Starting from the diode threshold voltage, the MR response of the devices starts increasing rapidly, as shown previously [10]. When the device current is less than or close to 1 μA, the MR traces are quite sharp and above a certain magnetic field value (typically 50 mT) the MR trace tends towards saturation.

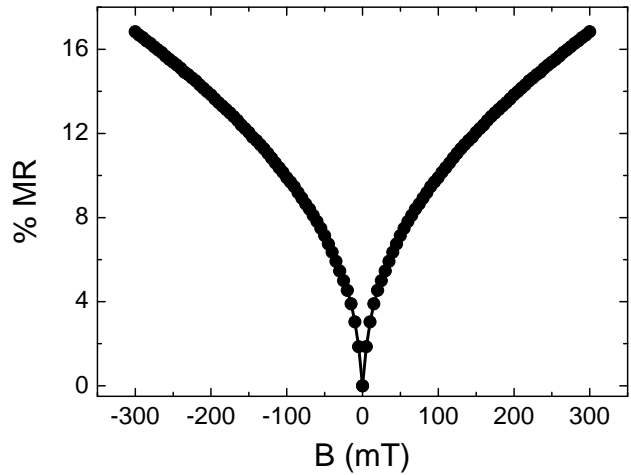

*Figure 1 Magnetoresistance response as a function of magnetic field (B) of a typical spin-coated RRP3HT diode at room temperature measured with 100 μA current.*

However, at higher current values, the MR line-shape starts changing clearly and saturation is no longer observed in the MR traces. As shown in Fig. 2, when the device current is above 1 µA, the complete magnetic field range (0 – 300 mT) cannot be fitted to the theoretically predicted Lorentzian ($\Delta I(B)/I \propto B^2/(B^2+B_0^2)$) or specific non-Lorentzian ($\Delta I(B)/I \propto B^2/(|B|+B_0)^2$ ) line shapes anymore (black line). The MR response can be fitted with a Lorentzian (red, dotted line) only at lower magnetic fields.

Above a certain magnetic field ($B_0$) the MR response shows a power law dependence of $B^{0.5}$ (blue, dashed line). Earlier, √B dependence of MR at higher B values has been observed in different inorganic disordered electronic systems and its origin was interpreted as an electron-electron interaction effect [14]. The two distinct line shapes at low and high magnetic fields indicates that the total MR response that we observe is the sum of multiple effects.

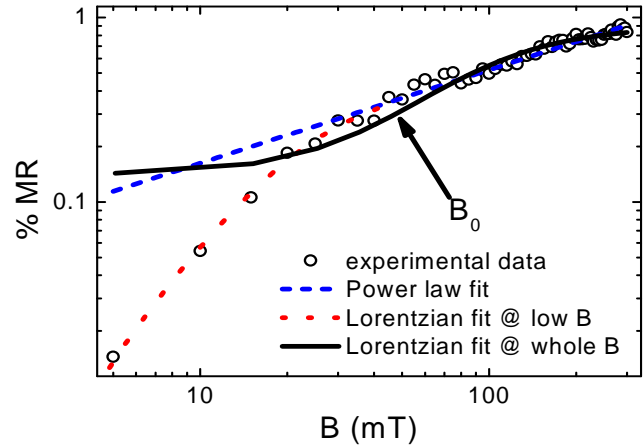

*Figure 2 Experimental MR data of a typical spin-coated RRP3HT diode at 300 K measured with 1 µA current and fitting of the experimental data with the Lorentzian fit for the whole range of B (black line) and for lower B values (red line) and with the power law fit (blue line).*

We further investigated the effect of different current densities on the MR properties. We, interestingly, observed that when the measuring current is increased the MR line shapes deviated more from the Lorentzian or non-Lorentzian line shapes. Fig. 3 shows MR response for a bias current of 100 µA. The $B_0$ value shifts towards even lower values when the current is higher. When the $B_0$ value is plotted as a function of different measuring current, it is observed that $B_0$ decreases exponentially as a function of the measuring current at room temperature (Fig. 4).

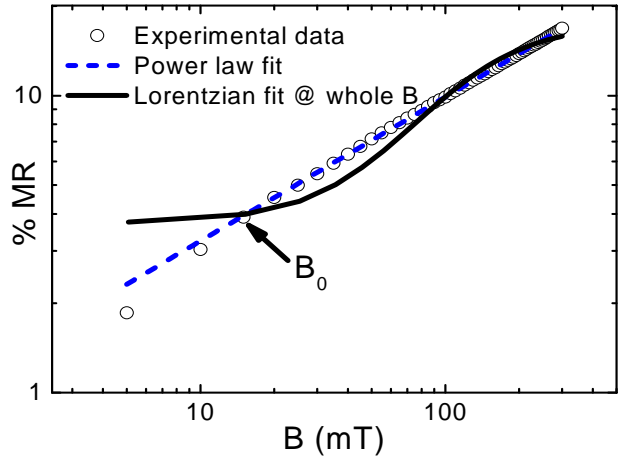

*Figure 3 Experimental MR data of a typical spin-coated RRP3HT diode at room temperature measured with 100 µA current and fitting of the experimental data with the power law and Lorentzian fits show clear deviation of $B_0$ to the low field side for measurements with a higher current.*

From these observations we can conclude that there are two distinct effects governing the MR response. At lower current densities and lower B values, the Lorentzian line shape indicates MR due to spin splitting whereas with increasing measuring current and higher B, the power law dependence of MR suggest inelastic scattering of charge carriers leading to increased positive MR values. Exponential dependence (showed by the fitting line in Fig. 4) of $B_0$ value with the measuring current indicates that as the current density is sharply increased in the diodes in their operating region, the spin splitting effect is no more important and only the scattering mechanism between the carriers control the MR response.

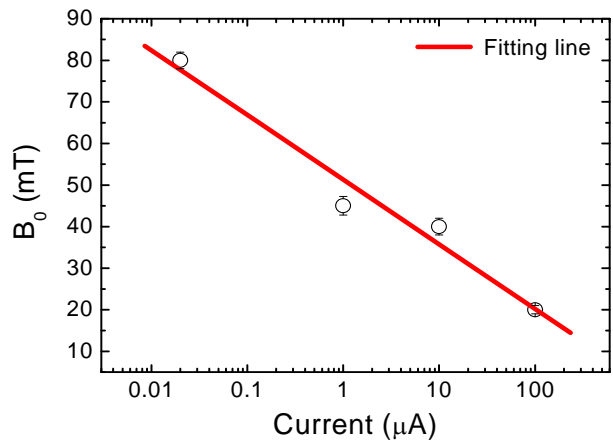

*Figure 4 $B_0$ as a function of measuring current show an exponential decrease with increasing current at 300 K.*

While the fundamental physics of the OMAR effect still remain debatable, the very high MR response in almost all OS materials has given rise to an interest in possible sensor applications. The emergence of printed electronics has also opened up the possibility of using such devices for low-end applications. This motivated us towards the prospect of inkjet printed organic diodes as magnetic sensors.

Fig. 5 shows the MR response of a typical inkjet printed diode measured with 1 µA current at room temperature. The devices showed positive MR response between 10 – 15%. Although the MR response of these devices are almost as good as the spin coated devices, the device yield is still less than 50% and also the devices are not as stable as the spin coated ones.

Among the difficulties with the printed devices, we found the spreading of the RRP3HT solution on the substrate, as well as some observed "coffee-stain effects" [15]. The problem with the "coffee-stain effect", i.e. uneven drying on the substrate, of the inkjetted semiconductor could probably be overcome by increasing the concentration (and viscosity) and by changing the solvents.

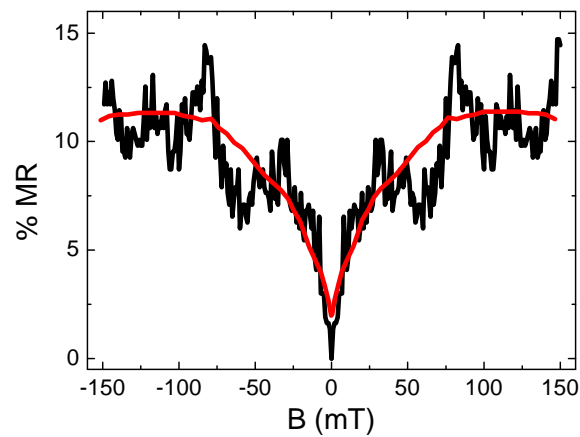

*Figure 5 %MR response of an inkjet printed diode measured with 1 µA current at room temperature. The red line is the guide to the eye.*

Another problem during inkjetting was the difference in wetting behaviour on the Al and the polyester substrate (and the P3HT surface) when printing the P3HT solution and on the P3HT surface and the polyester substrate when printing the Ag-nanoparticle dispersion. Better wetting on the P3HT surface would have been achieved e.g. by a surface plasma treatment, but could possibly also have detrimental effects on the diode characteristics.

In conclusion, we have observed OMAR effect in polymeric diodes fabricated both by using the conventional spin coating technique and by inkjet printing. High enough MR was recorded in both types of devices, which is extremely promising for immediate application as roll-to-roll printed flexible magnetic sensors but there is still huge scope for improvement. MR lineshapes are discussed as a function of the measuring current and different components of MR response was pointed out. Clearer physical understanding of the OMAR phenomenon and better control of printing methods are required.

**Acknowledgements**

The authors thank Harri Aarnio for the fabrication of some of the spin coated diodes. The authors also gratefully acknowledge the Academy of Finland for funding from project 116995. S.M. acknowledges the SITRA - CIMO fellowship.